\documentstyle[aps,epsf,preprint]{revtex}
\begin{document}
\def\n{{\bf\hat n}}
\def\dot{\!\cdot\!}
\def\cross{\!\times\!}
\def\grad{\nabla_{\scriptscriptstyle \perp}}
\title{Order and Frustration in Chiral Liquid Crystals}
\author{Randall D. Kamien}
\address{Department of Physics and Astronomy, University of Pennsylvania,
 Philadelphia, PA  19104, USA}
\author{Jonathan V. Selinger}
\address{Center for Bio/Molecular Science and Engineering, Naval Research
Laboratory, Code 6900, Washington, DC  20375, USA}
\date{\today}
\maketitle

\begin{abstract}
This paper reviews the complex ordered structures induced by chirality in
liquid crystals.  In general, chirality favors a twist in the orientation of
liquid-crystal molecules.  In some cases, as in the cholesteric phase, this
favored twist can be achieved without any defects.  More often, the favored
twist competes with applied electric or magnetic fields or with geometric
constraints, leading to frustration.  In response to this frustration, the
system develops ordered structures with periodic arrays of defects.  The
simplest example of such a structure is the lattice of domains and domain walls
in a cholesteric phase under a magnetic field.  More complex examples include
defect structures formed in two-dimensional films of chiral liquid crystals.
The same considerations of chirality and defects apply to three-dimensional
structures, such as the twist-grain-boundary and moir\'e phases.

\end{abstract}
\pacs{PACS numbers: }

\section{Introduction}

Frustration is the competition between different influences on a physical
system that favor incompatible ground states.  The phenomenon of frustration
has been studied in a wide range of systems because it makes nature
interesting:  Under competing influences, a system can develop structures with
complex spatial organization, and can have a rich variety of transitions
between different ordered phases.

A prototypical example of frustration on a lattice is an Ising antiferromagnet.
On a periodic, one-dimensional
lattice with an {\sl odd} number of sites $N$, the system
cannot have a perfect alternation in the spin, and hence it must have one
frustrated bond.  As a result, there are $2N$ degenerate ground states
corresponding to the number of locations for the frustrated bond.  The
situation becomes more complicated on a two-dimensional triangular lattice, in
which frustration leads to complex ordered structures with dimerization of the
spins~\cite{antiferromagnet}.  If there were no frustration between the
geometry and the interactions,
the rich phase diagram of these simple lattice systems would disappear.

Frustration can also occur in continuum systems.  For example, a superconductor
in a magnetic field either expels magnetic field in the superconducting phase
or becomes a normal metal and allows magnetic flux into the bulk.  The
Abrikosov vortex is the compromise that balances these frustrated extremes.
While the normal metal and the superconducting phases are translationally
invariant, the Abrikosov phase of a type-II superconductor is a flux-line
lattice, which creates a new level of organization at a new length scale set by
the flux-line interactions and the applied magnetic field~\cite{ChaLub}.

In this review, we discuss the frustration that occurs in the
liquid-crystalline phases of {\sl chiral} molecules.  A molecule is chiral if
it cannot be superimposed on its mirror image via any proper rotation or
translation.  Since the birth of stereochemistry in 1848 with Pasteur's
discovery of molecular handedness~\cite{PASTEUR}, the role of chirality has
become ever more important.  Pharmaceuticals, food additives, electro-optic
devices, and liquid crystal displays are examples of technologies that use or
rely on the effect of chiral constituents.

In the context of liquid crystals, the main effect of chirality is that chiral
molecules do not tend to pack parallel to their neighbors, but rather at a
slight skew angle with respect to their neighbors.  This packing can be
visualized, for example, by the packing of hard screws, although the details
of the chiral interaction are somewhat more subtle~\cite{HKLi,HKLii}.  As a
result, chirality favors a macroscopic twist in the orientation of the
molecules, with a characteristic chiral length scale.  This favored twist leads
to the self-assembly of periodically ordered structures.  Recently, these
self-assembling structures have been used as templates~\cite{rs} in the
nano- and
micro-fabrication of a host of technologically interesting
materials~\cite{PINE,Chaikin}, often with remarkable electro-optic properties.
Thus, the prediction, characterization, and control of self-assembling
structures is likely to be an essential ingredient in the quest for ever more
useful and economical devices.

In some cases, the twist in the molecular orientation favored by chirality can
be achieved without the introduction of any defects.  Two common examples of
such defect-free structures are the cholesteric phase and the helically twisted
smectic-C* phase.  More often, however, the twist favored by chirality competes
with other influences on a liquid-crystal system, such as applied magnetic or
electric fields or constraints imposed by geometry.  In those situations, the
competition leads to frustration, which causes the liquid crystal to develop
complex ordered structures.  Such structures are the subject of this paper.

The plan of this paper is as follows.  We begin with a cholesteric liquid
crystal in an applied magnetic field.  This example serves as a tutorial in
frustration because the mathematics is straightforward and can be solved
exactly.  The system is frustrated because of the competition between
chirality, which favors a twist in the molecular orientation, and the magnetic
field, which favors alignment of the molecules.  This frustration leads to the
formation of a series of domains in which the molecules are approximately
aligned with the field, separated by domain walls, or solitons, in which the
molecular orientation twists rapidly.

In the subsequent section, we consider defect structures that form in
two-dimensional films of liquid crystals.  These films have a different type of
frustration, which is not imposed by an applied field but is intrinsic to the
two-dimensional geometry:  Chirality favors a modulation in the molecular
orientation, but this modulation cannot be achieved in two dimensions without
the introduction of defects, {\sl i.e.} solitons in the molecular orientation.
We
discuss the ordered defect lattices that can arise, as well as variations in
the structures involving chiral symmetry-breaking and curvature in the film.

Finally, we consider defect structures in three-dimensional liquid-crystal
systems.  These structures include the twist-grain-boundary (TGB) phases, which
arises from the frustration of chirality combined with a smectic density wave.
Although these structures are much more complex than the cholesteric and
two-dimensional examples, they arise from the same considerations of order,
frustration, and defects.

\section{The Cholesteric in a Magnetic Field:  A Tutorial Example of
Frustration}

As an example of how chirality can frustrate liquid-crystalline order, we will
review the classic work of Meyer describing a cholesteric phase in a magnetic
field~\cite{Meyer}.  In this system, frustration arises from the competition
between chirality, which favors a twist in the molecular orientation, and
the applied magnetic field, which favors alignment of the molecules along the
field.  We will show that the system balances these opposing influences by
introducing domain walls -- one-dimensional analogues of Abrikosov vortices.
Indeed, domain walls and Abrikosov vortices can be regarded as different types
of solitons.

A cholesteric phase is described by the local director $\n$, a unit vector that
points along the average molecular direction.  The free energy of spatial
distortions in the director under a uniform magnetic field $\bf H$ is
given by the Frank free energy~\cite{Frank}
\begin{equation}
F = \frac{1}{2}\int d^3\!x\, \left\{K_1\left[\nabla\dot\n\right]^2
+K_2\left[\n\dot\nabla\cross\n + q_0\right]^2
+K_3\left[\n\cross\left(\nabla\cross\n\right)\right]^2
-\chi_{\scriptscriptstyle M}\left[{\bf H}\dot\n\right]^2\right\},
\label{frank}
\end{equation}
where $K_1$, $K_2$ and $K_3$ are elastic constants describing splay, twist, and
bend, respectively, $2\pi/q_0$ is the cholesteric pitch and
$\chi_{\scriptscriptstyle M}$ is the magnetic susceptibility of the director
(which we will take to be positive).

If the chiral parameter $q_0=0$, the system has a simple {\sl uniform} ground
state.  In this case, the director $\n$ is aligned with the magnetic field
$\bf H$, with no spatial variation.  By comparison, if the field ${\bf H}=0$,
the system has a simple {\sl twisted} ground state, with the continuously
twisting, defect-free cholesteric texture
$\n =\left[0,\cos(q_0 x),\sin(q_0 x)\right]$.  (This solution can, of course,
be rotated so that the pitch axis points in an arbitrary direction.)  The
question is now:  What happens if both $q_0$ and $\bf H$ are nonzero?

The simplest, approximate way to answer this question is to compare the free
energies of the uniform and twisted states.  The free energy density of the
uniform state in the presence of nonzero $q_0$ is
${\cal F}_{\rm uni}=\frac{1}{2}K_2\left[q_0^2 - \xi^{-2}\right]$, where
$\xi^{-2}\equiv \chi_{\scriptscriptstyle M}H^2/K_2$.  The free energy density
of the twisted state in the presence of nonzero $\bf H$, perpendicular to the
pitch axis, is ${\cal F}_{\rm twist} = -\frac{1}{4}K_2\xi^{-2}$. We can
compare these free energy densities to estimate the critical value of $q_0$ at
which one state is favored over the other:  we expect the twisted state
whenever $q_0\ge 1/(\sqrt{2}\xi)$.  However, this is only an approximation
because the system has many degrees of freedom and it can compromise between
the uniform and twisted states.  We will see that this estimate is close but
not correct.

For an exact solution, we consider a cholesteric phase with pitch along the
$x$-axis in a uniform magnetic field along the $z$-axis.  The director depends
on the angle $\theta(x)$: $\n=\left[0,\cos\theta(x),\sin\theta(x)\right]$. The
full free energy is
\begin{equation}
{F\over A} = \frac{1}{2}\int dx\, \left\{K_2\left[{d\theta(x)\over dx} -
q_0\right]^2 - \chi_{\scriptscriptstyle M}H^2\sin^2\theta(x)\right\},
\label{singordon}
\end{equation}
where we have divided out the $yz$ cross-sectional area $A$.  We recognize this
free energy as the classic Sine-Gordon model \cite{talapov}.   It can be
rewritten as the sum:
\begin{equation}
{F\over A} ={1\over 2}K_2\int dx\,
\left\{\left[\left({d\theta\over dx}\right)^2 + \xi^{-2}\cos^2\theta\right]
- 2q_0{d\theta\over dx} + q_0^2 \right\}
\label{bps}
\end{equation}
This decomposition enables us to study the effect of $q_0$ for fixed field.
The first pair of terms (in square brackets) favors uniform domains with
$\theta=(m+\frac{1}{2})\pi$, for any integer $m$, which all represent alignment
of the director along $\bf H$.  The next term favors an increase of
$\theta(x)$ from one domain to the next, giving a domain wall or soliton.  (The
final term is an unimportant constant.)  When $q_0$ becomes sufficiently large,
the system gains more free energy from a domain wall (from the $d\theta/dx$
term) than it loses (from the first pair of terms), and hence a single domain
wall forms.  As $q_0$ is increased above this threshold value, more domain
walls form, giving a periodic lattice of alternating domains and domain walls,
as shown in Fig.~1.  In the limit of high $q_0$, the density of domain walls
increases and the system approaches the continuously twisting cholesteric
state.

To find the threshold at which the first domain wall forms, the standard
approach is to minimize the free energy of Eq.~(\ref{bps}) by solving the
corresponding Euler-Lagrange equation~\cite{solution}.  Here, we will follow an
alternative approach, employing the method of Bogomol'nyi~\cite{Bogomolnyi},
which simplifies the calculation.  To do this, we rewrite Eq.~(\ref{bps}) as
\begin{equation}
{F\over A} ={1\over 2}K_2\int dx\,
\left\{\left[{d\theta\over dx} - \xi^{-1}\cos\theta\right]^2
- 2{d\over dx}\left[q_0\theta -\xi^{-1}\sin\theta\right]
+ \left[q_0^2 - \xi^{-2}\right]\right\} .
\label{bps2}
\end{equation}
The first term of this expression is positive-definite, and it vanishes if
\begin{equation}
\frac{d\theta}{dx}=\xi^{-1}\cos\theta .
\end{equation}
This differential equation has solutions corresponding to a uniform state
($\theta=\pm\pi/2$) and to a single domain wall
($\theta(x)=\pi/2-2\tan^{-1}[\exp(-(x-x_{\rm wall})/\xi)]$), and hence the
first term vanishes for both cases.  The difference in free energies between
these states therefore comes only from the second term -- a total
derivative which integrates to the boundaries at infinity and hence only
depends on the topological winding of the solution.  In the uniform state
$\theta(x)$ is the same at $\pm\infty$, but if there is a single domain wall
then $\theta(x)$ changes from $-\pi/2$ to $\pi/2$.  Hence, the free energy of a
single domain wall, compared with the free energy of the uniform state, is
\begin{equation}
{F_{\rm wall}-F_{\rm uni}\over A} = K_2\left(2\xi^{-1}-\pi q_0 \right).
\end{equation}
Thus, the single domain wall is favored for $q_0 \ge 2/(\pi\xi)$, close
to but smaller than our earlier estimate $q_0\ge 1/(\sqrt{2}\xi)$.

This example shows that a uniform ground state pointing along the magnetic
field is incompatible with chirality.  The frustration is resolved through the
introduction of a domain wall, or soliton, which takes the system from one
domain to another, equivalent domain.  In the following sections we will
describe higher-dimensional systems that share this feature -- the frustrating
effect of chirality will be relieved through solitons that connect equivalent
domains.  Although the mathematics will become more complex, the essential idea
will be the same.  In some cases these solitons will be {\sl topological
defects}~\cite{Mermin}, although we will not be focusing on that aspect in this
review.

\section{Defect Structures in Two Dimensions}

\subsection{Geometric Frustration}

In Sec.~II, we showed how an aligning magnetic field can be frustrated by
chiral interactions in materials.  This type of frustration is somewhat
artificial since the magnetic field is externally applied.  However, geometry
often plays a role similar to an external field -- boundary conditions in a
finite-sized sample often compete with the intrinsic
interactions~\cite{HexTh,HexEx}.  Similarly, when liquid crystals are confined
between two plates~\cite{Clinici}, in a capillary~\cite{Zumer,Capillary}, or in
a droplet~\cite{LubWei}, the shape of the sample can alter the bulk ground
state structures.

Geometric effects are even more profound in two-dimensional systems.  When
liquid crystals are confined to two dimensions, the system cannot have a
defect-free chiral modulation, as in a three-dimensional cholesteric phase in
zero magnetic field.  Rather, the system can only express structural
chirality by introducing defects, {\sl i.e.} solitons.

This geometric frustration can be seen graphically in Fig.~2.  This figure
shows a two-dimensional film of liquid crystals in a smectic-C or other tilted
phase.  The arrows represent the projection ${\bf c}(x,y)$ of the
three-dimensional molecular director $\n (x,y)$ into the smectic layer plane.
If the system were to have a defect-free chiral modulation, analogous to a
three-dimensional cholesteric phase, then $\bf c$ would have to rotate through
an angle of $2\pi$ as a function of $x$, as shown in the figure.  Is such a
structure favored by chirality?  Note that this structure can be divided into
alternating regions labeled 1 and 2.  These two regions are actually
{\sl mirror images} of each other.  For that reason, they cannot both be
favored by chirality.  If one is favored, then the other is disfavored to
exactly the same extent.  Hence, this structure will not occur in a
two-dimensional film of chiral liquid crystals.  Instead, the film must have
regions with the favored modulation in $\bf c$ (either 1 or 2, depending on the
handedness of the material), separated by defect walls, in which $\bf c$ jumps
back so that it can have the favored modulation once again.

The same geometric frustration can also be understood mathematically.  Again
consider a two-dimensional film of liquid crystals in the smectic-C phase.  If
the magnitude of the molecular tilt is uniform, so that the projection $\bf c$
has constant length, then the only chiral term in the bulk free energy is
\begin{equation}
F_{\rm chiral} = \int d^2 x [-\lambda {\bf\hat z}\cdot\nabla\times{\bf c}],
\end{equation}
which favors bend in $\bf c$. This term is a total derivative, which can be
reduced to a line integral around the edges of the liquid-crystal domain.
Moreover, if the modulation is in the $x$ direction, as in Fig.~2, the chiral
term simplifies to
\begin{equation}
F_{\rm chiral} = \int d^2 x \left[-\lambda
\frac{\partial c_y}{\partial x}\right],
\end{equation}
and the corresponding line integral is just
\begin{equation}
F_{\rm chiral} = -\lambda L [c_y(x_{\rm max}) - c_y(x_{\rm min})],
\end{equation}
where $L$ is the system size in the $y$ direction.  This expression for
the chiral term shows explicitly that chirality favors an increase in $c_y$
across the width of a domain, from $x_{\rm min}$ to $x_{\rm max}$.  The
magnitude of this increase is limited, because $c_y$ can at most increase from
$-1$ to $+1$.  Hence, if the system had only a single domain, it could have at
most a chiral free energy of $F_{\rm chiral} = -2\lambda L$, which does not
even scale with the system size in the $x$ direction.  The only way for the
system to have a more favorable chiral free energy is to break up into finite
domains separated by domain walls, giving a chiral contribution of roughly
$-2\lambda L$ for {\sl each domain}.  This tendency to form domain walls,
which can also be regarded as defect lines or solitons, is an expression of
geometric frustration.

As an aside, we note that a similar type of geometric frustration occurs in
two-dimensional films with an {\sl up-down} asymmetry rather than a chiral
asymmetry.  Such systems include Langmuir monolayers, Langmuir-Blodgett films,
and surface layers in liquid crystals.  In pioneering experimental and
theoretical work, Meyer and Pershan showed that the up-down asymmetry favors a
splay in $\bf c$~\cite{MeyerPershan}.  This is analogous to the bend favored by
chirality, if $\bf c$ is rotated by $90^\circ$, transforming bend into splay.
Hence, the results of the following sections apply also to films with up-down
asymmetry, with this rotation in $\bf c$.

\subsection{Lattice of Defect Lines}

We have argued in the previous section that two-dimensional films of chiral
liquid crystals have a geometric frustration, which leads them to break up into
domains separated by defect lines.  For that reason, one would expect such
films to show a lattice of defect lines.  Such a defect lattice has been
observed in polarization micrographs of freely suspended thin
films~\cite{clark,dierker}, and related but more complex lattices have been
seen in thicker films~\cite{gorecka}.  The defect lattice has been studied
theoretically by several authors~\cite{langer,hinshaw1,hinshaw2,jacobs}.  In
this and the following section, we discuss the theory of the defect lattice,
combining the theoretical approaches of those authors.

For the simplest and most macroscopic theory, we can regard the liquid-crystal
film as a sequence of alternating domains and domain walls.  In this view, the
domains can be described by continuum elastic theory, while the domain walls
are just defect lines with some phenomenological energy $\epsilon$ per unit
length.  For the domains, the free energy is just the chiral term discussed in
the previous section plus the two-dimensional Frank free energy,
\begin{equation}
F = \int d^2 x [-\lambda {\bf\hat z}\cdot\nabla\times{\bf c}
+\frac{1}{2}K_1(\nabla\cdot{\bf c})^2
+\frac{1}{2}K_3({\bf\hat z}\cdot\nabla\times{\bf c})^2],
\label{fstripe}
\end{equation}
A precise minimization of this free energy was discussed in Ref.~\cite{langer}.
However, the most important features of the results can be seen in the
following approximate calculation.  Suppose that the system has a striped
pattern as shown in Fig.~3, with a stripe width of $d$.  Across each stripe,
$\bf c$ rotates through an angle of order $\pi$, which gives
\begin{equation}
\nabla\cdot{\bf c}\approx
{\bf\hat z}\cdot\nabla\times{\bf c}\approx
\frac{1}{d}.
\end{equation}
Inserting these approximations into Eq.~(\ref{fstripe}) gives the free energy
of a single stripe,
\begin{equation}
F_{\rm stripe} = d L \left[-\frac{\lambda}{d}+\frac{\bar K}{d^2}\right],
\end{equation}
where $\bar K=\frac{1}{2}(K_1+K_3)$ is the mean Frank constant.  To this
expression must be added the free energy of a single domain wall,
\begin{equation}
F_{\rm wall} = L\epsilon.
\end{equation}
Combining these terms gives the free energy per unit area,
\begin{equation}
\frac{F}{A}=\frac{F_{\rm stripe}+F_{\rm wall}}{dL}=
-\frac{\lambda-\epsilon}{d}+\frac{\bar K}{d^2}.
\end{equation}
Minimizing this free energy density over $d$ gives the stripe width
\begin{equation}
d=\frac{2\bar K}{\lambda-\epsilon}.
\end{equation}

>From these results, we can draw several conclusions about the defect lattice.
First, the defect lattice only exists when the chiral coefficient $\lambda$
exceeds the defect line energy $\epsilon$.  This is reasonable, because the
sequence of alternating domains and domain walls is only favored in comparison
with the uniform state if the system is ``sufficiently chiral,'' that is, if it
gains more free energy from the modulation across a domain than it loses by
introducing a domain wall.  Increasing the chiral coefficient $\lambda$ reduces
the stripe width, because it favors the formation of more stripes.  By
contrast, increasing the defect line energy $\epsilon$ or the mean Frank
constant $\bar K$ increases the stripe width.  All three of these parameters
may be functions of system variables, such as temperature.  If changing
temperature causes $\lambda$ to pass through the value of $\epsilon$, then the
stripe width will diverge and the system will undergo a second-order transition
from the striped phase to the uniform phase.

The striped phase is not the only type of defect lattice that can form in
two-dimensional films of chiral liquid crystals.  Another theoretical
possibility, pointed out in Ref.~\cite{hinshaw1}, is the hexagonal lattice
shown in Fig.~4.  This lattice consists of hexagonal domains separated by
domain walls.  Each hexagonal domain has the favored chiral variation of
$\bf c$, and the domain walls allow this variation to be repeated periodically.
There are topological vortices in $\bf c$ at the centers of the domains, and
corresponding antivortices at the corners.  The full phase diagram for the
uniform phase, the striped phase, and the hexagonal lattice has not yet been
worked out.  However, we can say that the energy balance between the striped
phase and the hexagonal lattice depends on two factors.  First, the striped
phase has a combination of bend and splay in $\bf c$, while the hexagonal
lattice has more nearly pure bend.  Hence, the hexagonal lattice is favored in
a system with $K_1 \gg K_3$.  Second, the free energy of the hexagonal lattice
includes a logarithmic interaction between vortices and antivortices, in
addition to the terms discussed for the striped phase.  Hence, the hexagonal
lattice is favored in highly chiral systems, in which the lattice spacing
becomes small and this interaction becomes large and favorable.  To the best of
our knowledge, the hexagonal lattice has not yet been seen experimentally,
which suggests that these conditions have not been achieved.  It remains a
theoretical possibility for future experiments.

\subsection{More Microscopic View of Defect Lines}

In the previous section, we took a {\sl macroscopic} point of view, in which
the domains of the liquid-crystal film are described by continuum elastic
theory and the domain walls are just defect lines.  In this point of view, we
neglect the internal structure of the domain walls, and just suppose that the
domain walls have some energy $\epsilon$ per unit length.  This wall energy is
not predicted by the theory; rather, it is an input parameter for the theory.

It is possible to discuss the same lattice of domains and domain walls using a
more {\sl microscopic} point of view, in which the same theory describes both
the domains and the domain walls.  Such a microscopic theory is analogous to
the theory for a cholesteric phase in a magnetic field presented in Sec.~II, in
which the free energy functional was used both for the domains and for the
solitons between the domains.  The advantage of the more microscopic approach
is that it gives a model for the internal structure of the domain walls, and
makes a prediction for the domain wall energy.  The disadvantage is that the
microscopic theory is less general than the macroscopic theory -- a different
microscopic theory is needed for each type of domain wall.

One microscopic theory for the lattice of domains and domain walls was
developed in Refs.~\cite{hinshaw1,hinshaw2,jacobs}.  This theory supposes that
the magnitude of the molecular tilt in the smectic-C phase is not fixed, but
rather can vary as a function of position.  The tilt magnitude has a favored
value that it will assume in most of the film (the domains), but it can deviate
from this favored value in certain regions (the domain walls).  To describe
this variation mathematically, we allow the projection ${\bf c}(x,y)$ to vary
in length as well as direction.  The free energy then becomes
\begin{equation}
F=\int d^2 x
\left[-\frac{1}{2}r|{\bf c}|^2
+\frac{1}{4}u|{\bf c}|^4
-\lambda|{\bf c}|^2{\bf\hat{z}}\cdot{\bf\nabla}\times{\bf c}
+\frac{1}{2}K_1({\bf\nabla}\cdot{\bf c})^2
+\frac{1}{2}K_3({\bf\nabla}\times{\bf c})^2\right].
\label{SmCfreeenergy}
\end{equation}
Here, the $r$ and $u$ terms are the standard Ginzburg-Landau expansion of the
free energy in powers of $\bf c$.  They favor the tilt magnitude
$|{\bf c}|=\sqrt{r/u}$ in most of the film, but allow a different tilt
magnitude in certain regions.  The $\lambda$ term gives the favored variation
in the director due to molecular chirality.  This term is written as
$|{\bf c}|^2 {\bf\hat{z}}\cdot{\bf\nabla}\times{\bf c}$ rather than just
${\bf\hat{z}}\cdot{\bf\nabla}\times{\bf c}$ because the latter term is a total
divergence, which integrates to a constant depending only on the boundary
conditions.  By contrast,
$|{\bf c}|^2 {\bf\hat{z}}\cdot{\bf\nabla}\times{\bf c}$ is not a total
divergence because the factor of $|{\bf c}|^2$ couples variations in the
magnitude of $\bf c$ with variations in the orientation.

The minimization of this free energy was worked out in
Refs.~\cite{hinshaw1,hinshaw2,jacobs}.  Those studies show explicitly that the
optimum texture of ${\bf c}(x,y)$ breaks up into domains and domain walls.  In
the domains, the magnitude of $\bf c$ is approximately $\sqrt{r/u}$, and the
orientation of $\bf c$ bends in the sense favored by chirality.  In the domain
walls, the magnitude of $\bf c$ is greatly reduced, and the orientation bends
back in the opposite sense.  The system does not lose as much free energy from
the reverse bend in the domain walls as it gains from the forward bend in the
domains because of the coupling between variations in the magnitude of $\bf c$
with variations in the orientation.  As in the macroscopic theory, the domains
and domain walls can be arranged in stripes or in a hexagonal lattice.

An alternative microscopic theory has been developed specifically for
{\sl tilted hexatic} phases, which are known as smectic-I, -F, and
-L~\cite{dierker} (see also related work~\cite{SelingerHexatic,SelingerMRS}).
These phases have order in the orientation of the molecular tilt {\sl and}
order in the orientations of the intermolecular ``bonds'' (not chemical bonds,
but lines indicating the directions from one molecule to its nearest neighbors
in the smectic layer).  These two orientations are coupled by an interaction
potential that favors a particular alignment of the molecular tilt with respect
to the bond directions.  The favored alignment is along a nearest-neighbor bond
direction in the smectic-I phase, halfway between two nearest-neighbor bonds in
the smectic-F phase, or at an intermediate orientation in the smectic-L phase.
In a tilted hexatic phase of {\sl chiral} molecules, shown in Fig.~5, the
molecular chirality causes the tilt direction to rotate across a domain, and
the bond direction must rotate with it in order to keep the favored alignment.
In a domain wall, the tilt direction jumps by $60^\circ$ from one bond
direction to another, equivalent bond direction.  Hence, the tilt and bond
directions are locked together everywhere in the domains, and only deviate from
the favored alignment in the domain walls.  The line energy and width of the
domain walls can both be calculated in terms of the tilt-bond interaction
potential and the elastic constants~\cite{SelingerHexatic,SelingerMRS}.  Hence,
this theory gives a specific model for the relationship between the tilt and
bond directions in the domain walls, but it reduces to the same
macroscopic theory
studied earlier for the chiral texture of the domains.

The domain walls might have other internal structures that have not yet been
considered in the literature.  For example, in a two-component mixture of
liquid crystals, the domains might have the optimal composition of the two
components, while the domain walls would have a different composition.
Equivalently,
in a system with a single liquid-crystal component plus impurities, the domain
walls might be regions where the impurities preferentially accumulate.  All
that is necessary is that is the domain walls must have a {\it different}
microscopic structure than the domains, and that the different microscopic
structure must reduce the free-energy cost for short-length-scale variations in
the tilt direction $\bf c$.  Any such microscopic structure can give the
macroscopic structure discussed in the previous section, with a periodic
lattice of domains and domain walls.

\subsection{Smectic-A Phase under Electric Field}

The work discussed in Secs.~III.A--C is concerned with two-dimensional films of
liquid crystals in phases with spontaneous tilt order, {\sl i.e.} the smectic-C
or
tilted hexatic phases.  A recent paper has shown that related considerations
apply to films of the untilted smectic-A phase of chiral molecules under an
electric field~\cite{SelingerElecField}.  In the absence of an electric field,
the smectic-A phase does not have tilt order; the average molecular director is
aligned along the smectic layer normal.  This director alignment along the
layer normal is unaffected by chirality (unless the chirality is so strong that
it actually disrupts the smectic layer structure, as discussed in Sec.~IV
below).  When an electric field is applied in the smectic layer plane, it
induces a tilt of the director away from the layer normal.  This induced tilt
is called the electroclinic effect.  The system then experiences a frustration
analogous to the cholesteric phase in a magnetic field:  the field favors a
particular alignment of the induced tilt, while chirality favors a modulation
in the tilt direction.

The smectic-A phase under an electric field has been modeled in
Ref.~\cite{SelingerElecField} using the free energy
\begin{eqnarray}
F=\int d^2 x
\biggl[&&\frac{1}{2}r|{\bf c}|^2
+\frac{1}{4}u|{\bf c}|^4
+b{\bf\hat{z}}\cdot{\bf E}\times{\bf c}
-\lambda|{\bf c}|^2{\bf\hat{z}}\cdot{\bf\nabla}\times{\bf c}\nonumber\\
&&+\frac{1}{2}K_1({\bf\nabla}\cdot{\bf c})^2
+\frac{1}{2}K_3({\bf\nabla}\times{\bf c})^2\biggr].
\end{eqnarray}
This free energy is equivalent to Eq.~(\ref{SmCfreeenergy}) for the smectic-C
phase, except that the coefficient of $|{\bf c}|^2$ is positive, indicating
that the smectic-A phase does not have spontaneous tilt order, and a coupling
of the tilt to the applied field $\bf E$ has been added.  The question is
whether the minimum of this free energy is a uniform tilt or a chiral
modulation in ${\bf c}(x,y)$.  This was investigated through a combination of
continuum elastic theory and lattice simulations.  The results showed that the
state of uniform tilt can become unstable to the formation of chiral stripes,
similar to the stripes seen in films of the smectic-C phase, which resolve the
frustration between chirality and the applied field.  The theoretical phase
diagram, shown in Fig.~6, predicts that stripes will occur for a certain range
of field.  This range becomes larger as the chiral coefficient $\lambda$
increases and as the system approaches the transition from smectic-A to
smectic-C, where $r$ passes through 0.

In most electroclinic liquid crystals developed for device applications, this
modulation has not been seen, suggesting that the chiral coefficient $\lambda$
is not large enough to give stripes.  However, certain electroclinic liquid
crystals show an otherwise unexplained striped modulation that is consistent
with this chiral mechanism~\cite{SelingerElecField,bartoli,sprunt}.  Hence,
controlling this effect may become important for optimizing liquid-crystal
devices.

\subsection{Chiral Symmetry-Breaking}

So far we have considered the effects of chirality on systems of fixed
chirality.  Some related effects can occur in systems that undergo a chiral
symmetry-breaking transition, in which they spontaneously break reflection
symmetry and select a handedness.  Some possible mechanisms for such
symmetry-breaking were proposed in Ref.~\cite{swbk}.  The first and simplest is
just phase separation of a racemic mixture.  If a mixture of opposite
enantiomers separates into its chiral components, then any local region of the
system is predominantly right- or left-handed.  An appropriate chiral order
parameter would then be the difference in densities of the two enantiomers.  A
second possibility, which can occur even in systems of pure achiral molecules,
is that the molecules can pack in a two-dimensional film in two inequivalent
ways that are mirror images of each other.  The chiral order parameter would
then be the difference of densities of the two types of packing.  A third
possibility is the formation of the tilted hexatic phase known as smectic-L,
mentioned in Sec.~III.C, which breaks reflection symmetry in the relationship
between the tilt and bond directions.

If a two-dimensional liquid-crystal film undergoes a chiral symmetry-breaking
transition, how does the spontaneous chirality affect the ordering of the film?
This question has been addressed in Refs.~\cite{swbk,ohyama,iwamoto}; here we
review the discussion in the first of those papers.  The free energy can be
written in the form
\begin{equation}
F=\int d^2 x \left[
{1\over2}\kappa(\nabla\psi)^2
+{1\over2}t\psi^2+{1\over4}u\psi^4
+{1\over2}K_1(\nabla\cdot{\bf c})^2
+{1\over2}K_3(\nabla\times{\bf c})^2
-\lambda\psi\nabla\times{\bf c}\right].
\end{equation}
Here, $\psi(x,y)$ is the pseudoscalar order parameter representing the extent
and handedness of chiral symmetry-breaking, and ${\bf c}(x,y)$ is the
projection of the molecular director into the layer plane.  (This study assumes
the tilt is constant in magnitude; Ref.~\cite{ohyama} allows it to vary.)  The
first three terms in $F$ are a Ginzburg-Landau expansion in powers of $\psi$,
the next two terms are the Frank free energy for variations in $\bf c$, and the
final term couples these variables.  Note that $\nabla\times{\bf c}$ is
multiplied by a chiral order parameter which can itself vary across the film.

By minimizing the free energy over $\psi(x,y)$ and ${\bf c}(x,y)$,
Ref.~\cite{swbk} obtained the phase diagram shown in Fig.~7.  At high
temperature, the system is in a uniform nonchiral phase.  As the temperature
decreases, the system undergoes a continuous chiral symmetry-breaking
transition.  Because the local order parameter $\psi(x,y)$ is nonzero, it
favors a bend of the director, just as in the chiral films discussed in
Secs.~III.A--D.  In this case, however, since the system has spontaneous rather
than fixed chirality, it has a different way to resolve the geometric
frustration associated with chirality.  Instead of introducing domain walls in
which the director jumps back rapidly, this system can simply reverse the sign
of $\psi(x,y)$, which favors a continuous backward bend of the director.
Hence, the system breaks up into stripes of alternating right- and left-handed
chirality, as shown in Fig.~8.  At high temperature, the stripes involve a
smooth, sinusoidal modulation of both $\psi$ and $\bf c$.  As the temperature
decreases, this smooth modulation crosses over into a sharper modulation, with
domains of approximately constant positive or negative $\psi$ separated by
domain walls.  These domain walls, unlike the domain walls studied in systems
of fixed chirality, are solitons in which $\psi$ changes sign but $\bf c$ is
continuous.  (At high chirality, the system can also have a square lattice or
``checkerboard'' phase, with alternating square cells of right- and left-handed
chirality.)  At low temperature, the spacing of the soliton walls diverges and
the system has a transition into a phase of uniform chirality.  This phase may
or may not show the stripes studied in Secs.~III.B--C, with fixed $\psi$ and
solitons in $\bf c$, depending on the free-energy cost of such solitons.

The clearest experimental demonstration of this effect has been in experiments
on freely suspended liquid-crystal films~\cite{pang}.  These experiments show a
transition from a uniform nonchiral phase to a striped phase, which has a
spontaneous bend in $\bf c$.  The sign of the bend $\nabla\times{\bf c}$
alternates in successive stripes.  The stripe width diverges at the phase
transition, in at least qualitative agreement with the theoretical prediction.

\subsection{Spiral Defects}

For the striped phases discussed in the previous sections, the lowest-energy
state is to have straight, parallel stripes.  However, thermal fluctuations can
lead to curvature of the stripes.  Moreover, {\sl defects} in the striped
phases can have an interesting spiral form, as shown in Fig.~9(a).  Such spiral
defects have been seen experimentally in freely suspended films of chiral
liquid crystals in tilted hexatic phases~\cite{dierker}.  They have also been
seen in Langmuir monolayers of achiral molecules~\cite{knobler,praefcke} --
systems in which the stripes are driven by the up-down asymmetry rather than
the chiral asymmetry, as discussed at the end of Sec.~III.A.

The formation of spiral defects has been attributed to the following
mechanism~\cite{SelingerSpiral}.  Suppose there is a point vortex in $\bf c$.
A point vortex could arise from a localized impurity, from the kinetics of
formation of the monolayer, from boundary conditions on a circular droplet, or
from thermal fluctuations that nucleate a vortex-antivortex pair.  Near the
vortex core, there is more than the optimal bend.  Away from the vortex core,
the bend $\nabla\times{\bf c}$ decreases as $1/r$, where $r$ is the distance
from the core.  As a result, the defect generates domain walls, like the walls
in the periodic striped pattern, and thereby increases the bend up to the
optimal value.  Farther from the vortex core, the bend continues to decrease,
or equivalently, the distance between the domain walls increases linearly with
$r$.  To maintain the optimal bend, the pattern of domain walls may buckle to
form a right- or left-handed spiral, which maintains a constant spacing between
walls, as in Fig.~9(a).  Alternatively, the system may continue to generate
more domain walls in a dense branching morphology, as shown in Fig.~9(b).
Calculations have shown that the spiral has a lower free energy than the dense
branching morphology for large system size, and hence it is the more common
defect form.

Although it is tempting to associate the spiral form of the defect with the
chirality of the molecules, one should note this mechanism depends only on the
combination of a striped phase with a point vortex in $\bf c$.  It can occur in
films with an up-down asymmetry driving splay stripes, as well as in films with
chirality driving bend stripes.  The resulting spirals can be either right- or
left-handed.  Hence, the handedness of these defects can be regarded as a
{\sl macroscopic} chiral symmetry-breaking on the length scale of the defect,
in contrast with the microscopic chiral symmetry-breaking discussed in the
previous section.

\subsection{Membrane Curvature:  Ripples and Tubules}

In the discussion so far, we have seen how {\sl flat} liquid-crystal films
respond to the geometric frustration caused by chirality.  But do the films
need to remain flat?  That depends on the type of film.  Freely suspended films
of thermotropic liquid crystals are generally stretched across an aperture, and
hence are constrained by surface tension to remain nearly flat.  Langmuir
monolayers and Langmuir-Blodgett films also must remain nearly flat because
they must conform to the shape of the substrate.  However, lyotropic liquid
crystals, consisting of lipid membranes separated by solvent, have much more
freedom to curve out of the plane.  For those systems, membrane curvature is
another possible response to chirality.

Recent papers have considered chirality as a mechanism for inducing the
$P_{\beta'}$ ``ripple'' phases that are observed in lipid
membranes~\cite{ripple1,ripple2,ripple3}.  These studies show that any
modulation in the tilt director $\bf c$ will induce a modulation in the height
of the membrane above a flat reference plane, thus giving a periodic ripple in
the membrane curvature.  Several distinct ripple phases with different symmetry
are possible, depending on the orientation of $\bf c$ with respect to the
ripple wavevector.  Chirality is important for the structure of these phases
because it leads to a coupling between tilt variations and curvature that is
not present in nonchiral systems, and this coupling leads to asymmetric ripple
shapes that are consistent with experimental results.

Lipid membranes are not limited to slight ripple distortions about a basically
flat structure; they can also assume quite different shapes.  As an important
example, diacetylenic lipids form cylindrical tubules, with a typical diameter
of 0.5~$\mu$m and a typical length of 10~$\mu$m to 1~mm~\cite{SchnurScience}.
These tubules have been studied extensively for use as templates for the
formation of metal cylinders, for applications in electroactive composites and
controlled-release systems.  The formation of lipid tubules has been explained
theoretically as a response to the chirality of the lipid molecules.  A
continuum elastic theory for flexible chiral membranes has shown that there is
an intrinsic bending force on any chiral bilayer membrane with tilt
order~\cite{HelfrichProst}.  This early work has been extended in several ways
by different investigators.  In particular,
Refs.~\cite{SelingerTubule1,SelingerTubule2} have shown that chirality has two
simultaneous effects on a membrane:  it induces the formation of a cylinder,
with a radius that can be calculated in terms of the continuum elastic
coefficients, and it induces the formation of stripes in the orientation of the
molecular tilt on the cylinder.  These stripes are equivalent to the stripes in
flat films of chiral liquid crystals discussed in Secs.~III.B--C.  They wind
around the cylinder in a helical fashion, as shown in Fig.~10, and they should
also induce slight ripples in the cylindrical curvature. Theoretical work on
tubules and related lipid microstructures is reviewed and
compared with experiments in a forthcoming publication~\cite{TubuleReview}.

\section{Defect Structures in Three Dimensions}

\subsection{The Twist-Grain-Boundary Phase}

In three dimensions, competing intrinsic interactions can lead to complex
defect structures.  A prototypical example of this is the twist-grain-boundary
phase (TGB) of chiral, smectogenic liquid crystals, the analogue of the
Abrikosov vortex lattice phase of type-II superconductors \cite{TGBreview}.  In
this system, the periodic smectic order is incompatible with the helical
cholesteric order favored by the chiral molecules.  As in two dimensions, we
can either appeal to a more macroscopic picture in which defects allow for
regions of twist and assume an energy penalty per unit length, or we can employ
a microscopic picture which includes the relevant order parameter and
correlation lengths to model the defect structure.

In this example we will use a microscopic picture, due to the (relative)
simplicity of the chiral smectic-$A$ phase.  To model this system one
introduces a complex order parameter $\psi$.  The magnitude of $\psi$ is a
measure of smectic order, while the phase of $\psi$ is associated with the
spontaneously broken translational invariance of the layered smectic phase.
The density of material is simply $\rho(x) = \rho_0 + {\rm Re} \psi(x)$
where $\rho_0$
is the background density.  In the smectic-A phase the layer normal is parallel
to the local director $\n$.  When the molecules are chiral, the director
undergoes fluctuations controlled by the free energy of Eq.~(\ref{frank}) with
magnetic field ${\bf H}=0$.  In the smectic phase we may write
$\psi = \vert\psi\vert \exp\{iq\left(z-u(x,y,z)\right)\}$, where we have chosen
the average layer normal to point along $\n_0 =\hat z$ and where $u(x,y,z)$ is
the displacement field that describes fluctuations in the layered structure.
To quadratic order in $\delta\n = \n - \n_0$ and $u$, the free energy of this
system is~\cite{ChaLub}:
\begin{equation}
F = \frac{1}{2}\int d^3\!x\, \left\{B(\partial_z u)^2 +
\tilde B(\grad u -\delta\n)^2 + K_1(\grad\dot\delta\n)^2
+K_2(\grad\cross\delta\n+q_0)^2 + K_3(\partial_z\delta\n)^2\right\} ,
\label{London}
\end{equation}
where $B$ and $\tilde B$ are elastic constants proportional to
$\vert\psi\vert^2$.  This free energy, which is similar to the London free
energy of a superconductor, is frustrated.  It is possible to choose solutions
$u$ and $\delta\n$ which are independent of $z$ and for which
$\grad\dot\delta\n=0$, and so the effective free energy is simply
\begin{equation}
F=\frac{1}{2} \int d^3\!x\, \left\{\tilde B(\grad u -\delta\n)^2 +
K_2(\grad\cross\delta\n + q_0)^2\right\}.
\label{Londonii}
\end{equation}
Note that this free energy, which is the sum of two positive definite terms,
can {\sl never vanish}:  if the first term in Eq.~(\ref{Londonii}) vanishes
then $\grad u =\delta\n$, so $\grad\cross\delta\n=\grad\cross\grad u$ must
vanish identically and the second term is nonzero.  Conversely, if the second
term vanishes then $\grad\cross\delta\n = -q_0$ and there is no nonsingular
solution $u$ that can make the first term vanish.  Indeed, if
$\grad\dot\;\delta\n =0$ then $\delta\n = q_0y\hat x$ and it is clear that the
first term in Eq.~(\ref{Londonii}) will diverge with system size so strongly
that even the free energy density will diverge.  As in the case of the
cholesteric in the magnetic field, however, we can identify two extreme
solutions.  As we have argued, if $\tilde B$ is nonzero then we must choose
$\grad\cross\delta\n=0$ for a finite energy density.  The smectic phase thus
completely expels twist.  Another possibility, however, is to lose the smectic
order:  since $B$ is proportional to $\vert\psi\vert^2$ it is possible to
destroy the smectic order and consequently have $B$ vanish.  Then
$\grad\cross\delta\n = -q_0$ and the cholesteric phase ensues.

Not surprisingly, just as in the example in Sec.~II, it is possible to have a
mixture of both effects.  By letting $\vert\psi\vert$ vanish in isolated
regions it is possible to relieve the intrinsic frustration in this system.  It
is straightforward to check~\cite{ChaLub} that an extremum of
Eq.~(\ref{Londonii}) is
\begin{eqnarray}
u&=&{md\phi\over 2\pi}\nonumber\\
\delta\n_\phi&=&{md\over 2\pi\lambda}{\cal K}_1(\rho/\lambda)
-{md\over 2\pi \rho}\\
\delta\n_\rho&=&0 \nonumber
\end{eqnarray}
where $m$ is an integer, $d$ is the smectic layer spacing, ${\cal K}_1(\cdot)$
is the modified Bessel function of order 1, $\rho=\sqrt{x^2+y^2}$,
$\phi=\tan^{-1}(y/x)$, and $\lambda\equiv\sqrt{K_2/\tilde B}$ is the twist
penetration depth.  A measure of the defect's strength is its Burgers vector
$b\equiv md$ -- an integer multiple of the layer spacing.  While $u$ is
singular at the origin, $\delta\n$ is well defined.  However, in order for the
free energy to remain finite, the bulk modulus $B$ must vanish at the origin,
and thus $\vert\psi\vert\rightarrow 0$ in a region of size $\xi$.

While the physics of a single defect (as shown in Fig.~11) was well understood
for many years~\cite{deGennes}, the full geometry resulting from a
proliferation of defects was determined by Renn and Lubenksy in
1988~\cite{TGBTh}.  It was discovered soon thereafter in the homologous series
$S-1-{\rm methyheptyl} 4'-[(4''-n-{\rm alkyloxyphenyl})
{\rm propionoyloxyl-biphenyl-4-carboxylate}]$ (nP1M7) later that
year~\cite{TGBEx}.  Since the TGB phases exist over narrow temperature ranges,
it is quite likely that they had been overlooked in prior studies -- it is in
this way that theoretical constructs serve as guideposts for experiment.  The
TGB phase is composed of a sequence of grain boundaries, separated by $\ell_b$,
with each of these boundaries being composed of screw dislocations $\ell_d$
apart, as shown in Fig.~12.  To understand how a periodic array of screw
dislocations effects a twist in the smectic layer normal, it is convenient to
employ the Poisson summation formula to determine the strains
$\partial_i u(x,y,z)$.  In Ref.~\cite{scherk} it was found that, for screw
dislocations parallel to the $z$-axis with Burgers vector $b$, separated by
$\ell_d$ along the $y$-axis:
\begin{eqnarray}
\partial_x u & = & -{b\over 2\ell_d} {\sin(2\pi y/\ell_d)\over \cosh(2\pi
x/\ell_d)
-\cos(2\pi y/\ell_d)},\nonumber\\
\partial_y u & = & {b\over 2\ell_d} {\sinh(2\pi x/\ell_d)\over \cosh(2\pi
x/\ell_d)
-\cos(2\pi y/\ell_d)}, \\
\partial_z u & = & 1- \sqrt{1-\left(b/2\ell_d\right)^2}.\nonumber
\label{strains}
\end{eqnarray}
By comparing the layer normals ${\bf N} = \left[\partial_x u,\partial_y u,
\partial_z u -1\right]/\sqrt{1-2\partial_zu +(\nabla u)^2}$ at $x=\pm\infty$,
one can check that the array of defects rotates the smectic structure by an
angle $\alpha = 2\sin^{-1}(b/2\ell_d)$.   The TGB phase is an assembly of such
grain boundaries.  Recent progress has been made on calculating the geometry of
the quiescent state of the TGB phase -- the ratio $\ell_b/\ell_d$ is relatively
constant over a range of parameters and is approximately $0.94$~\cite{bluv}.
This is both qualitatively and quantitatively in very close agreement with the
experiments of Navailles and coworkers~\cite{navailles}.

\subsection{Chiral Lyotropic Phases}

Since the TGB phase is a defect phase of a layered system, it is natural to ask
whether a single lamella might tear to form the analogue of screw
dislocation~\cite{ChiralLyotropic}.  Unlike the screw dislocation in a smectic,
however, the core of the defect is not nematic but is filled with solvent.
Thus, unlike the systems we have discussed so far, the core size is not set by
parameters in a Landau theory but by the balance between elasticity and the
line tension of an exposed edge.  Although reminiscent of the tubule phases
described in section II.G, these helicoidal structures are different in detail.
Recall that for a surface we may define two invariant measures of curvature:
The mean curvature is the average of the two principal curvatures while the
Gaussian curvature is the product.  In tubules the mean curvature couples to
the chirality, while in the helicoids the mean curvature vanishes.  Moreover,
when integrated over a closed surface the Gaussian curvature yields a
topological constant and is thus often neglected.  However, because the
helicoid is formed by tearing the plane, the total Gaussian curvature can
change.  This effect may play a role in promoting and stabilizing the newly
discovered smectic blue phases~\cite{Pansu,KamSm} as well.

\subsection{The Moir\'e Phase}

While the smectic liquid crystal represents one-dimensional translational
order, the hexagonal columnar phase is an example of two-dimensional
translational order in the plane perpendicular to the local director $\n$.  It
can be thought of as stacks of disks arranged in a hexagonal lattice.  However,
in each stack there is no translational order -- the disks form one-dimensional
liquids.  An elasticity theory can be developed for this phase~\cite{KNMoire}
based on the principles of broken symmetry.  Within this theory, the defect
line tension may be calculated and one finds that screw dislocations, as
depicted in Fig.~13, have a finite energy per unit length.  We will take a more
macroscopic approach to discussing these defect phases, using only the
finiteness of the line tension from the more microscopic theory.

Because the columnar crystal has two displacement or phonon modes, it is
possible to have more than one type of defect phase.  Two have been proposed,
although other defect complexions may be possible.  The first is a close
analogue of the smectic TGB phase: the polymer tilt-grain-boundary phase.  In
this phase, regions of undistorted hexagonal columnar crystal are separated by
grain boundaries which effect a twist of the local nematic director.  As in the
TGB phase, the cholesteric rotation is broken into discrete jumps.  The single
grain boundaries are each composed of an array of parallel screw dislocations.
Again, the combined effect of these defects may be calculated~\cite{KNMoire},
and they lead to a twisting of the columnar phase.

A different defect phase is also possible: the moir\'e phase.  This phase is,
in some sense, a three-dimensional analogue of the hexagonal domain phase
described in Sec.~II.B.  Here, however, the defect lattice rotates the
intermolecular bonds, twisting the crystal along the director axis as shown in
Fig.~14.  This phase can be thought of as a stack of hexagonal crystals, each
successive one rotated by a special angle to produce a ``moir\'e'' pattern.
Recent experiments on nucleosome core particles~\cite{Livolant} may exhibit
this exotic phase.

\section{Acknowledgments}
RDK was supported by NSF CAREER Grant DMR97-32963 and the Alfred P. Sloan
Foundation.  JVS was supported by the Naval Research Laboratory and the Office
of Naval Research.

\epsfclipon

\begin{figure}
\vbox{\centering\leavevmode\epsfxsize=3.375in\epsfbox{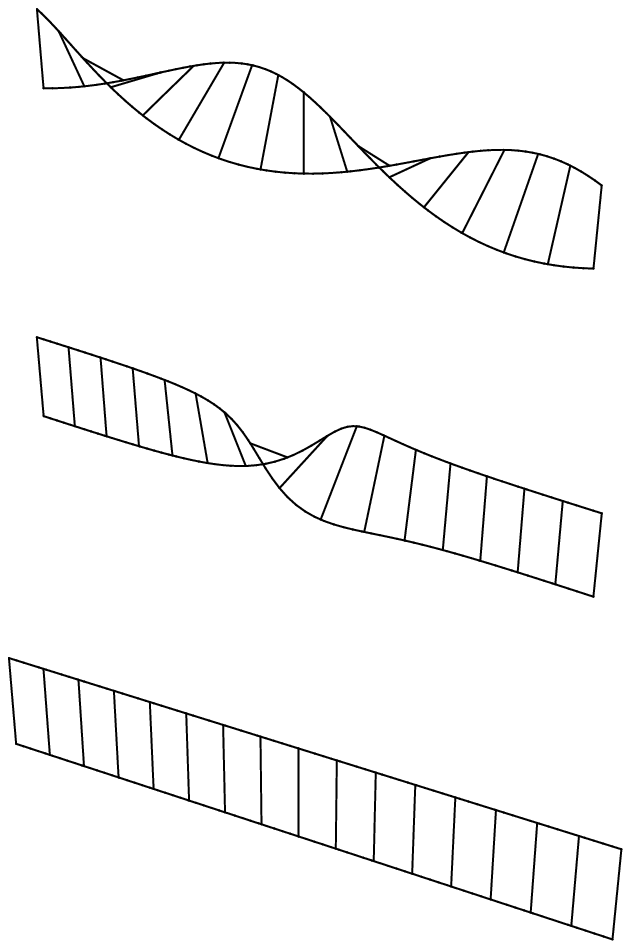}\bigskip
\caption{Cholesteric helix under an applied magnetic field.  Top:  Continuously
twisting helix, in the limit of $q_0\gg 2/(\pi\xi)$.  Middle:
Single domain wall
between two uniform domains, for $q_0\protect\gtrsim 2/(\pi\xi)$.  Bottom:
Uniform state, for $q_0 < 2/(\pi\xi)$.}}
\end{figure}

\begin{figure}
\vbox{\centering\leavevmode\epsfxsize=3.375in\epsfbox{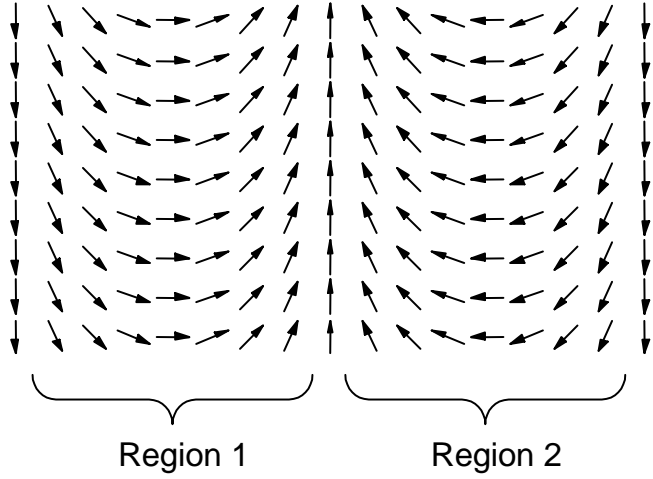}\bigskip
\caption{Hypothetical defect-free modulation in a two-dimensional film of
liquid-crystals in a smectic-C or other tilted phase.  The arrows represent the
projection of the three-dimensional molecular director into the smectic layer
plane.  Note that regions 1 and 2 are mirror images of each other, with
opposite signs of the bend in the director.  Hence, this structure is {\sl not}
favored by chirality.}}
\end{figure}

\begin{figure}
\vbox{\centering\leavevmode\epsfxsize=3.375in\epsfbox{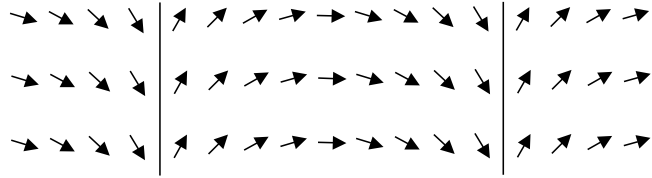}\bigskip
\caption{Striped defect lattice of alternating domains and domain walls in a
two-dimensional film.  Unlike the structure in Fig.~2, this pattern is favored
by chirality, because all the domains have the preferred chiral variation in
the director.  Adapted from Ref.~\protect\cite{hinshaw1}.}}
\end{figure}

\begin{figure}
\vbox{\centering\leavevmode\epsfxsize=3.375in\epsfbox{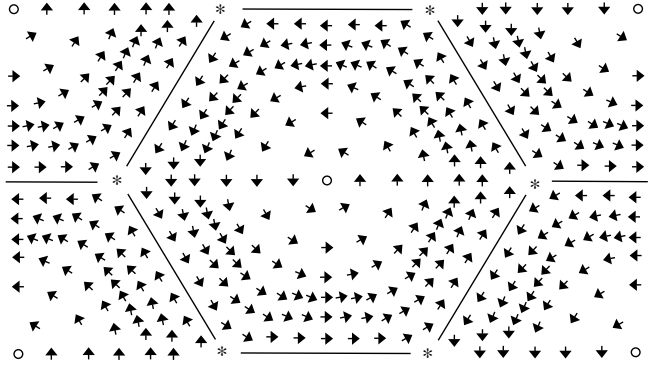}\bigskip
\caption{Hexagonal defect lattice of domains separated by domain walls.  Each
domain has the favored chiral variation of the director.  The structure has
vortices at the centers of the domains and antivortices at the corners.
Adapted from Ref.~\protect\cite{hinshaw1}.}}
\end{figure}

\begin{figure}
\vbox{\centering\leavevmode\epsfxsize=3.375in\epsfbox{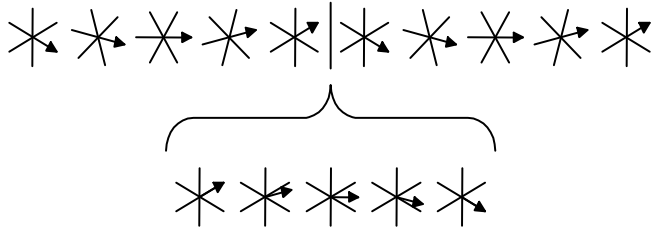}\bigskip
\caption{Striped defect lattice in a tilted hexatic (smectic-I) phase.  The
six-pointed stars represent the hexatic bond directions, and the arrows
represent the tilt direction ({\sl i.e.} the projection of the molecular
director
into the smectic layer plane).  The top half of the figure shows that the tilt
and bond directions rotate together across each domain.  The bottom half shows
that the tilt deviates from the favored alignment with the bonds inside the
narrow defect walls.}}
\end{figure}

\begin{figure}
\vbox{\centering\leavevmode\epsfxsize=3.375in\epsfbox{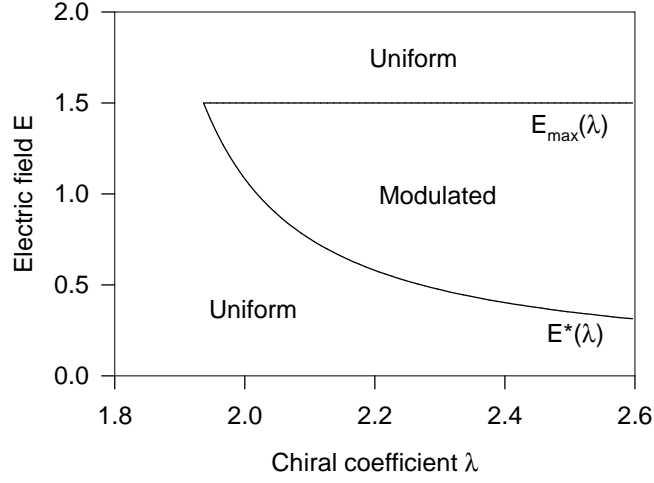}\bigskip
\caption{Phase diagram for a smectic-A film of chiral molecules under an
applied electric field, showing the uniform state and the chiral modulation.
The phase diagram is expressed in terms of the chiral coefficient $\lambda$ and
the applied field $\bf E$, for fixed parameters $r=0.5$, $u=1$, $b=1$, and
$\bar{K}=\frac{1}{2}(K_1 +K_3)=1.5$.  From
Ref.~\protect\cite{SelingerElecField}.}}
\end{figure}

\begin{figure}
\vbox{\centering\leavevmode\epsfxsize=3.375in\epsfbox{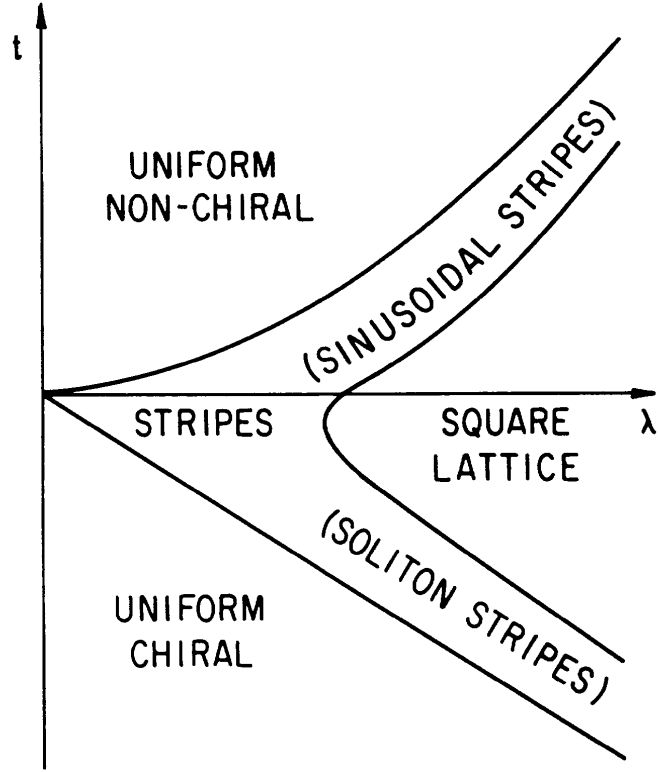}\bigskip
\caption{Mean-field phase diagram for chiral symmetry-breaking in a
two-dimensional film.  The parameter $t$ represents temperature, while
$\lambda$ is the coupling between the chiral order parameter $\psi$ and the
curl of the tilt director field $\bf c$.  This phase diagram is a schematic
view, not drawn to scale.  From Ref.~\protect\cite{swbk}.}}
\end{figure}

\begin{figure}
\vbox{\centering\leavevmode\epsfxsize=3.375in\epsfbox{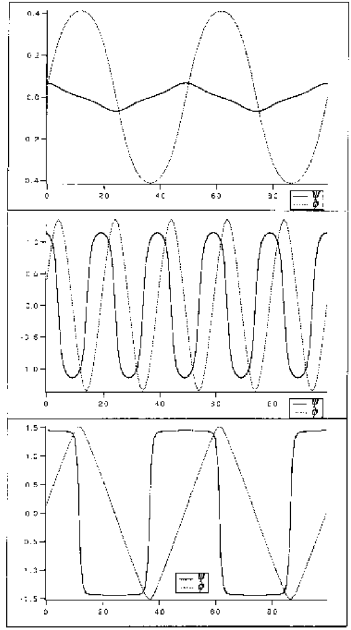}\bigskip
\caption{The modulation in the chiral order parameter $\psi(x)$ and the tilt
director ${\bf c}(x)=(\cos\phi(x),\sin\phi(x))$ at three temperatures in the
symmetry-breaking striped phase.  Top:  $t=0.9$.  Middle: $t=-1$. Bottom:
$t=-2$.  In all three plots, $\lambda=K_1=K_3=\kappa=u=1$. Note the evolution
from the sinusoidal-stripe regime at high temperature to the soliton-stripe
regime at low temperature.  From Ref.~\protect\cite{swbk}(b).}}
\end{figure}

\begin{figure}
\vbox{\centering\leavevmode\epsfxsize=3.375in\epsfbox{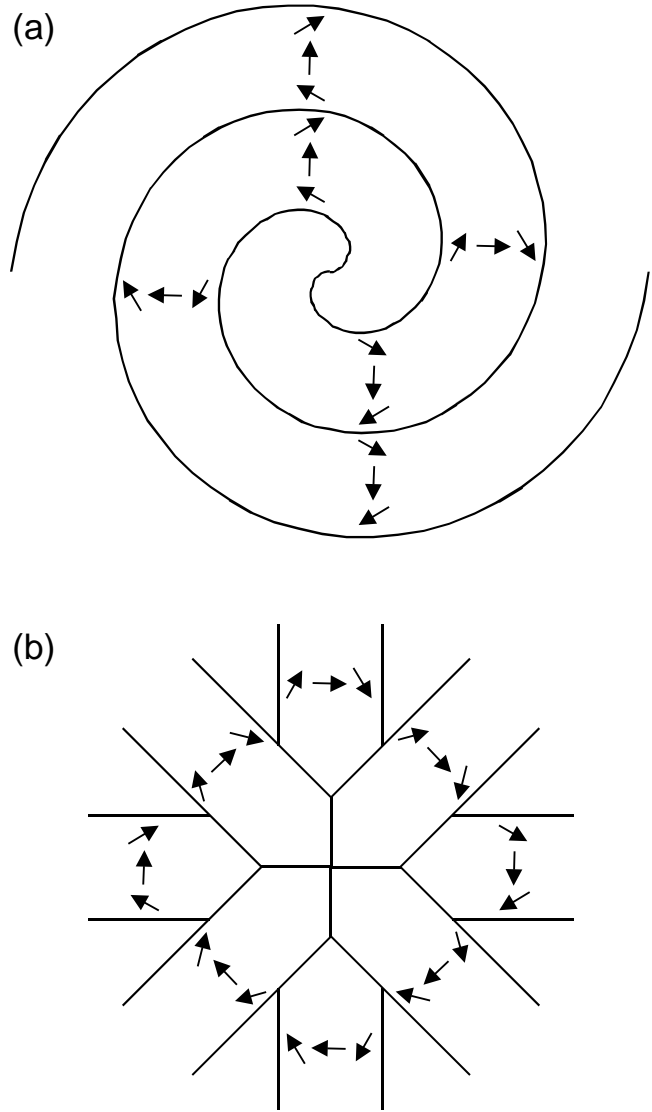}\bigskip
\caption{(a) Spiral defect resulting from the combination of a striped phase
with a point vortex in the tilt director $\bf c$.  (b) Dense branching
morphology, an alternative structure for a point vortex in a striped phase.
The spiral defect generally has a lower free energy than the dense branching
morphology.  Adapted from Ref.~\protect\cite{SelingerSpiral}.}}
\end{figure}

\begin{figure}
\vbox{\centering\leavevmode\epsfbox{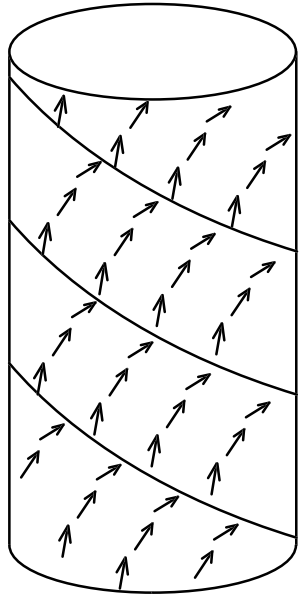}\bigskip
\caption{Striped pattern in the tilt director in the modulated state of a lipid
tubule.  The arrows indicate the direction of the molecular tilt, projected
into the local tangent plane.  From
Refs.~\protect\cite{SelingerTubule1,SelingerTubule2}.}}
\end{figure}

\begin{figure}
\vbox{\centering\leavevmode\epsfxsize=3.375in\epsfbox{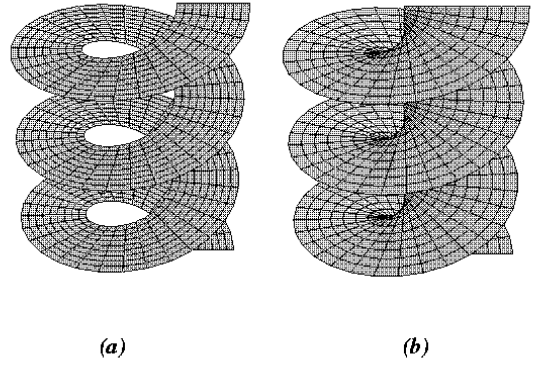}\bigskip
\caption{Structure of a single screw dislocation in a smectic-A phase. In
(a) the core is shown, which is typically nematic.  In (b)
there is no hole, appropriate for a description which ignores the
details of the core.}}
\end{figure}

\begin{figure}
\vbox{\centering\leavevmode\epsfxsize=3.375in\epsfbox{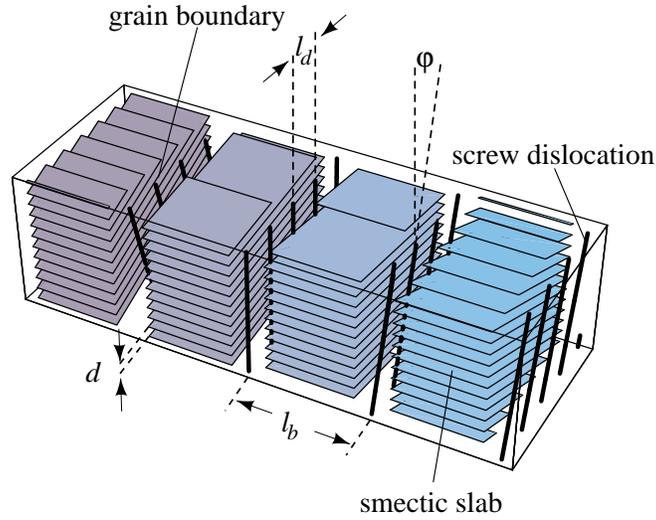}\bigskip
\caption{Twist-grain-boundary (TGB) phase, consisting of a smectic-A phase with
a sequence of grain boundaries separated by a distance $\ell_b$, with each of
these grain boundaries composed of screw dislocations separated by a distance
$\ell_d$. Adapted from Ref.~\protect\cite{TGBTh}.}}
\end{figure}

\begin{figure}
\vbox{\centering\leavevmode\epsfxsize=3.375in\epsfbox{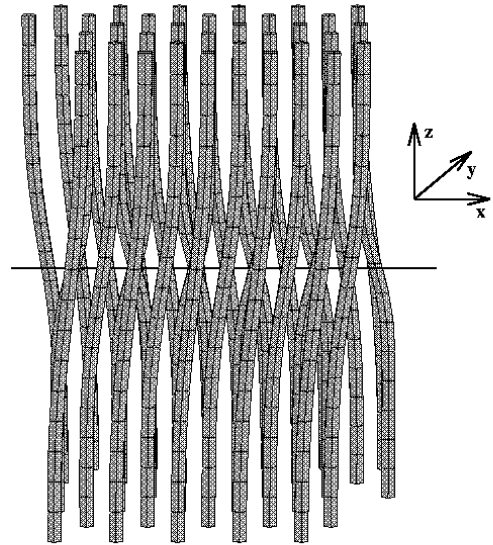}\bigskip
\caption{Structure of a single screw dislocation in a hexagonal columnar
phase.  From Ref.~\protect\cite{KNMoire}.}}
\end{figure}

\begin{figure}
\vbox{\centering\leavevmode\epsfxsize=3.375in\epsfbox{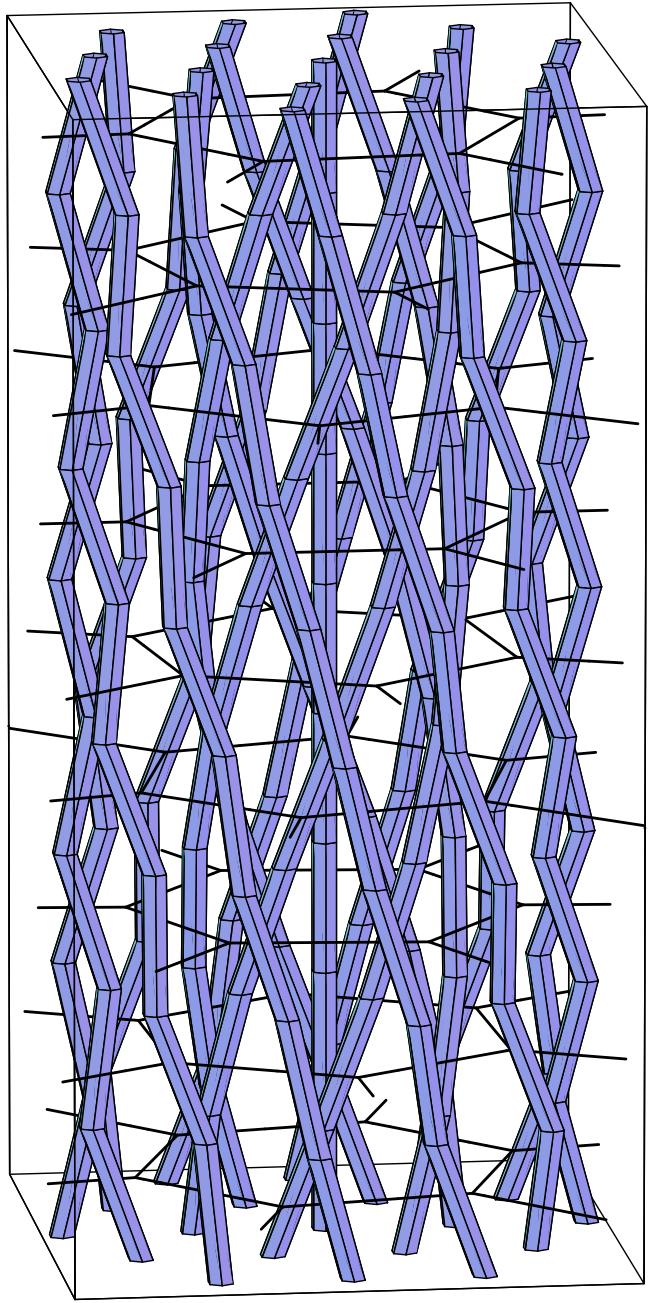}\bigskip
\caption{Moir\'e phase, composed of a lattice of screw dislocations in a
hexagonal columnar phase. From Ref.~\protect\cite{KNMoire}.}}
\end{figure}

\end{document}